\documentclass[bibyear]{aa}
\usepackage{amsmath}
\usepackage{graphicx}
\usepackage{txfonts}
\usepackage{amssymb}
\usepackage[]{natbib}
\usepackage{mathrsfs}
\usepackage{caption}
\usepackage{float}
\usepackage{afterpage}
\usepackage{amstext}

\newcommand{\wasp}[0]{\mbox{WASP-33}}
\newcommand{\waspb}[0]{\mbox{WASP-33~b}}

\begin{document}

   \title{A temperature inversion in WASP-33b?  
     \thanks{The LBT is an international collaboration among
       institutions in the United States, Italy and Germany. LBT
       Corporation partners are: The University of Arizona on behalf
       of the Arizona university system; Instituto Nazionale di
       Astrofisica, Italy; LBT Beteiligungsgesellschaft, Germany,
       representing the Max-Planck Society, the Leibniz-Institute for
       Astrophysics Potsdam, and Heidelberg University; The Ohio State
       University, and The Research Corporation, on behalf of The
       University of Notre Dame, University of Minnesota and
       University of Virginia.}}
   \subtitle{Large Binocular Telescope occultation data confirm
     significant thermal flux at short wavelengths}
   \author{C. von Essen$^{1,4}$, M. Mallonn$^2$, S. Albrecht$^1$,
     V. Antoci$^1$, A. M. S. Smith$^3$, S. Dreizler$^4$,
     K. G. Strassmeier$^2$} \authorrunning{C. von Essen et al. (2015)}
   \titlerunning{Secondary eclipse of WASP-33b}
   \offprints{cessen@phys.au.dk}

   \institute{$^1$Stellar Astrophysics Centre, Department of Physics and Astronomy, Aarhus University, Ny Munkegade 120, DK-8000 Aarhus C, Denmark\\ 
              $^2$Leibniz-Institut f\"ur Astrophysik Potsdam, An der Sternwarte 16, 14482, Potsdam, Germany\\
              $^3$N. Copernicus Astronomical Centre, Polish Academy of
Sciences, Bartycka 18, 00-716, Warsaw, Poland\\
              $^4$Institut f\"ur Astrophysik G\"ottingen, Friedrich-Hund-Platz 1, 33037, G\"ottingen, Germany\\
              \email{cessen@phys.au.dk}
               }
   \date{Received 27/03/2015; accepted 10/07/2015}
\abstract{We observed a secondary eclipse of
  \waspb\ quasi-simultaneously in the optical \mbox{($\sim$0.55
    $\mu$m)} and the near-infrared \mbox{($\sim$1.05 $\mu$m)} using
  the 2$\times$8.4~m Large Binocular Telescope. \wasp\ is a $\delta$
  Scuti star pulsating with periods comparable to the eclipse
  duration, making the determination of the eclipse depth
  challenging. We use previously determined oscillation frequencies to
  model and remove the pulsation signal from the light curves,
  isolating the secondary eclipse. The determined eclipse depth is
  \mbox{$\Delta$F = 1.03$\pm$0.34 parts per thousand}, corresponding
  to a brightness temperature of \mbox{T$_{\rm B}$ = 3398 $\pm$ 302
    K}. Combining previously published data with our new measurement
  we find the equilibrium temperature of \waspb\ to be \mbox{T$_{\rm
      B}$ = 3358 $\pm$ 165 K}. We compare all existing eclipse data to
  a blackbody spectrum, to a carbon-rich non-inverted model and to a
  solar composition model with an inverted temperature structure. We
  find that current available data on \waspb's atmosphere can be best
  represented by a simple blackbody emission, without the need for
  more sophisticated atmospheric models with temperature
  inversions. Although our data cannot rule out models with or
    without a temperature inversion, they do confirm a high brightness
    temperature for the planet at short wavelengths. \waspb\ is one
  of the hottest exoplanets known till date, and its equilibrium
  temperature is consistent with rapid reradiation of the absorbed
  stellar light and a low albedo.}

\keywords{stars: planetary systems -- stars: individual: WASP-33 --
  methods: observational}
          
   \maketitle

\section{Introduction}

Hot Jupiters offer us a great astrophysical laboratory for atmospheric
science under extreme conditions as their atmospheres reach
temperatures above \mbox{1000 K}. Secondary eclipse measurements
easily separate the stellar from the planetary lights and are
therefore usually used to probe exoplanet atmospheres. Such
measurements are most straight forward in the near infrared (NIR),
where the planet-to-star contrast ratio is large \citep[see
  e.g.,][]{Deming2007,Christiansen2010,Knutson2011}. The first
detection of secondary eclipses were made from the space using the
infrared Spitzer Space Telescope
\citep{Charbonneau2005,Deming2005}. Nonetheless, the planetary
atmospheric composition and temperature structure has also been
detected from the ground \citep[see
  e.g.,][]{Snellen2008,Bean2010}. Early studies suggest the existence
of a temperature inversion layer caused by strong stellar absorption
of certain molecules high up in the exoplanet atmospheres. For hot
Jupiters, theoretical predictions place TiO and VO as favorite
candidates \citep[see e.g.,][]{Hubeny2003}. However, this is still
under debate \citep{Spiegel2009}. \cite{Fortney2008} proposed a
differentiation of hot Jupiters into two populations divided by the
amount of stellar irradiation. For these two families, the temperature
profiles would be dominated by the state of TiO and VO. However, more
recent observations have not been able to clearly show the existence
of these two states \citep{Knutson2010,Crossfield2012}. A good example
is WASP-12b, for which no inversion has been yet detected
\citep{Crossfield2012} despite being one of the most irradiated and
hottest exoplanets known \citep[$T_{\rm eq}\sim$3000
  K,][]{Hebb2009,LopezMorales2010}. As observing techniques and data
reduction strategies improve with time, past claimed detection of
temperature inversions are being overruled. For example, \mbox{HD
  209458b} was observed using Spitzer Space Telescope by
\cite{Knutson2008}. The four bandpass observations were represented by
a thermal inversion in the exoplanet atmosphere \citep[see
  e.g.,][]{Line2014}. However, \cite{Diamond_Lowe2014} used state of
the art techniques and models to perform a complete and
self-consistent analysis of the data, finding no temperature inversion
after all. To date, there doesn't seem to be a clear determinant for
which planets should present (or not) temperature
inversions. Measuring the structure of one of the hottest exoplanet
atmospheres could help to clarify the situation.

This work focuses on the transiting hot Jupiter
\waspb\ \citep{CollierCameron2010}. With a wavelength-dependent
brightness temperature of \mbox{3620~K}, \waspb\ is one of the hottest
exoplanets known to date \citep{Smith2011}. \wasp\ \mbox{(HD\,15082)}
is a $\delta$ Scuti star, oscillating with amplitudes in the
milli-magnitude regime and periods that are comparable to the transit
duration \citep{Herrero2011}. \waspb\ orbits the star each $\sim$1.22
days in a retrograde orbit \citep{CollierCameron2010}. Showing an
unusually large radius, \waspb\ belongs to the class of anomalously
inflated exoplanets \citep[][]{CollierCameron2010}.

The first secondary eclipse was detected by \cite{Smith2011} at
\mbox{0.91 $\mu$m}. \cite{Deming2012} observed two events in the
$K_{\rm S}$ band \mbox{(2.15 $\mu$m)}, and one secondary eclipse each
at 3.6 $\mu$m and \mbox{4.5 $\mu$m} using Warm Spitzer. The authors
found the data to be consistent with two different atmospheric models,
the first one with an inverted temperature structure and a solar
composition, and the second one with a non-inverted temperature
structure and a carbon-rich composition. Furthermore,
\cite{deMooij2013} re-observed \waspb's eclipses in the $K_{\rm S}$,
finding consistency with \cite{Deming2012} and determined \waspb's
effective temperature to be around \mbox{3300 K}. Recently,
  \cite{Haynes2015} observed \wasp\ during two occultations using the
  Hubble Space Telescope between 1.13 and 1.63 $\mu$m. The authors
  represented their observations by a dayside atmosphere with a
  temperature inversion and an oxygen-rich composition with a slightly
  sub-solar abundance of H$_2$O. Excluding this work, previous
studies on \waspb's emission were observed in a broad wavelength
range. At that time the authors didn't count with a data-independent
pulsation model nor with the characterization of the pulsation
spectrum of the host star. In \cite{vonEssen2014} we presented a two
and a half years long follow-up campaign of \wasp. We collected 457
hours of in- and out- of transit data that were used to characterize
the pulsation spectrum of \wasp\ with unprecedented precision, finding
8 statistically significant frequencies with periods comparable to the
transit duration \mbox{($\sim$2.7 hours)}. In this work, we observed
\wasp\ during secondary eclipse at \mbox{1.05 $\mu$m} using the Large
Binocular Telescope. We make use of the identified pulsations to
``clean'' the photometry and determine the eclipse depth of \waspb's
secondary eclipse. In this work, Section~\ref{Obs_Data} presents the
observations and the data reduction, Section~\ref{Fitting} a detailed
description of the model parameters and fitting procedures,
Section~\ref{SecTrans} the secondary eclipse measurements in the
context of the atmospheric structure of \wasp, and Section~\ref{Concl}
our final remarks.

%%%%%%%%%%%%%%%%%%%%%%%%%%%%%%%%%%%%%%%%%%%%%%%%%%%%%%%%%%%%%%%%%%%%%%%
%%%%%%%%%%%%%%%%%%%%%%%%%%%%%%%%%%%%%%%%%%%%%%%%%%%%%%%%%%%%%%%%%%%%%%%

\section{Observations and Data Reduction}
\label{Obs_Data}

We observed \wasp\ on October 12/13, 2012, during a secondary eclipse
with the 2$\times$8.4~m Large Binocular
Telescope\footnote{\url{www.lbto.org}} (LBT) in binocular mode. In
particular, we employed the NIR spectrograph LUCI \citep{Seifert2010}
simultaneously to the Large Binocular
Cameras\footnote{\url{lbc.mporzio.astro.it}}
\citep[LBC][]{Giallongo2008}. Unfortunately, due to technical problems
the spectroscopic LUCI time-series lacks the pre-ingress part of the
light curve. Therefore, our work focuses on the photometric data only.

The LBC consists of two wide-field cameras mounted on the prime focus
swing arms of the LBT. Photometric data were acquired using the LBC
Red camera, which is optimized for observations between 0.55 and 1
$\mu$m. The camera has four charge coupled devices (CCDs), each one of
them with its own gain and read-out noise
characteristics\footnote{\url{http://abell.as.arizona.edu/~lbtsci/Instruments/LBC/}}. To
minimize read-out time we used sub-frames of the CCDs with the
brightest stars within the field of view centered on each chip. The
size of the final sub-frames were approximately \mbox{1$\times$3}
arcmin, which reduced the read-out time from 30 to 15 seconds.

During the night the observing conditions were stable and the sky
  was clear. Seeing varied between 0.4 and 1.1 arcsec. The
photometric data were acquired using the Johnson-Cousins $V$ filter
centered at \mbox{0.55 $\mu$m} and a $Y$ filter centered at \mbox{1.05
  $\mu$m}, the latter with a full-width at half-maximum of \mbox{0.2
  $\mu$m}. As \wasp\ is relatively bright \mbox{(V$\sim$8.3)}, we
defocused the telescope to avoid saturation. The transmission function
of the $Y$-band peaks around 95\%, but the quantum efficiency of the
CCD drops down to 10\% in the filters wavelength range. This was also
useful to avoid saturation. \wasp\ is located in a sparse stellar
field. Therefore, the defocusing technique did not produce overlapping
of the stellar point spread functions, except for the optical
companion identified by \cite{Moya2011}. A discussion on third-light
contribution is addressed in Section~\ref{lab:thirdLight}. Exposure
times were of 0.74 seconds for the $V$-band and 3.23 seconds for the
$Y$-band. Every 3 images the filters were switched between the two
bands. Since the readout time is shorter than the time that the filter
wheel requires to go from one filter to the other \mbox{($\sim$20
  sec)}, the sampling between consecutive images is each $\sim$18
seconds, while the sampling between consecutive groups of 3 images of
the same filter is each $\sim$2.3 minutes.

We used the IRAF's package {\it ccdproc} for bias subtraction and the
flat fielding, and the IRAF task {\it apphot} for aperture
photometry. We measured fluxes of \wasp\ and three reference stars,
namely \mbox{BD+36~ 493}, \mbox{BD+36~ 487}, and \mbox{BD+36~ 488},
using various aperture radii. We then fixed the aperture radius so
that the scatter of the differential light curve is minimized. The
chosen aperture radius was of 74 pixels, equivalently to 16 arcsec
considering the plate scale of the LBC Red \mbox{(0.2255
  "/pixel)}. Next we divided \wasp's measured flux by the average of
the three reference stars producing the final differential light
curves (see Figure~\ref{fig:LBT_V} and Figure~\ref{fig:LBT_Y}). Since
these are influenced by stellar pulsations, we estimated the scatter
averaging the standard deviations calculated within 15 minutes bins,
where the deformations caused by \wasp's intrinsic variability can be
neglected. The derived uncertainties are \mbox{$\sim$0.8
  parts-per-thousand (ppt)} for the $V$-band and \mbox{0.6 ppt} in the
$Y$-band, four and three times the expected photometric precision
  assuming only photon noise (0.2 ppt), respectively. We believe that
  the difference between the achieved photometric precision and the
  theoretical photon-noise precision is caused by scintillation noise
  due to the short exposure times. To ensure that the photometry is
based on proper references, the three reference stars were plotted
against each other to check for constancy.

During the observations, the LUCI (spectroscopic) side of the LBT was
the one guiding the telescope. The LBC side followed, however, its
pointing stability was rather poor. After 3.5 hours of observations
\wasp\ and the reference stars drifted $\sim$200 pixels in both x
  and y directions close to the edge of the read-out windows and flux
was lost. These $\sim$0.5 hour of observations were omitted from our
analysis. \waspb\ eclipse lasts 2.7 hours. Therefore, our observations
account for $\sim$25 minutes of off-eclipse data before eclipse begins
and ends, respectively, and full eclipse coverage. As a final step the
time stamps were converted from Julian Dates to Barycentric Julian
Dates (BJD$_{\rm TDB}$) using the tools made available by
\cite{Eastman2010}.

\section{Results}
\label{Fitting}

\subsection{Analysis of the optical light curve}
\label{sec:VDATA}

The data presented here comprise quasi-simultaneous observations
during secondary eclipse of \waspb\ around the $V$ and $Y$ bands. The
predicted planet-star flux ratio in the $V$-band is \mbox{0.2 ppt},
four times lower than the accuracy of our measurements. Therefore, we
can neglect the planet imprint and use this band to measure the
stellar pulsations, and most specifically to tune their current phases
\citep[see Phase Shifts in][]{vonEssen2014}. Particularly, our model
for the light contribution of the stellar pulsations consists of eight
sinusoidal pulsation frequencies with corresponding amplitudes and
phases. Hence, to reduce the number of 24 free parameters and the
values they can take, we use prior knowledge about the pulsation
spectrum of the star that was acquired during \cite{vonEssen2014}. As
the frequency resolution is 1/$\Delta$T \citep{Kurtz1983}, 3.5 hours
of data are not sufficient to determine the pulsations frequencies.
Therefore, during our fitting procedure we use the frequencies
determined in \cite{vonEssen2014} as starting values plus their
derived errors as Gaussian priors. As pointed out in
\cite{vonEssen2014}, we found clear evidences of pulsation phase
variability with a maximum change of \mbox{2$\times$10$^{-3}$ c/d}. In
other words, as an example after one year time a phase-constant model
would appear to have the correct shape with respect to the pulsation
pattern of the star, but shifted several minutes in time. To account
for this, the eight phases were considered as fitting parameters. The
\cite{vonEssen2014} photometric follow-up started in August, 2010, and
ended in October, 2012, coinciding with these LBT data. We then used
the phases determined in \cite{vonEssen2014} during our last observing
season as starting values, and we restricted them to the limiting
cases derived in Section~3.5 of \cite{vonEssen2014}, rather than
allowing them to take arbitrary values. The pulsation amplitudes in
$\delta$ Scuti stars are expected to be wavelength-dependent
\citep[see e.g.][]{DaszynskaDaszkiewicz2008}. Our follow-up campaign
and these data were acquired in the blue wavelength range. Therefore
the amplitudes estimated in \cite{vonEssen2014}, listed in
Table~\ref{tab:MODEL_PARAMS}, are used in this work as fixed
values. This approach would be incorrect if the pulsation amplitudes
would be significantly variable \citep[see
  e.g.,][]{Breger2005FGVir}. Nonetheless, the short time span of LBT
data, and the achieved photometric precision compared to the
intrinsically low values of \wasp's amplitudes, make the detection of
any amplitude variability impossible.

\begin{table*}[ht!]
  \centering
  \caption{\label{tab:MODEL_PARAMS} Comparison of derived pulsations
    between previous and current work. From left to right: Frequency
    number (PN), frequencies in c/d, amplitudes in ppt, phases in
    degrees and range of variability for the phases determined in
    \cite{vonEssen2014} and used here as model
    parameters. Additionally, derived frequencies and phases
    \mbox{($V$,$Y$)} found in this work and phase differences in
    degrees. Errors are in all cases at 1-$\sigma$ level.}

  \scalebox{0.9}{
  \begin{tabular}{c c c c c | c c c c c c}
     \hline \hline
PN    & $\nu$                &    A          &$\phi_{\rm V}$  &$\Delta\phi_{\rm V}$&  $\nu_{\rm V,fit}$&  $\phi_{\rm V,fit}$ &   $\nu_{\rm Y,fit}$  & $\phi_{\rm Y,fit}$ & $\phi_{\rm Y-V}$  \\
      & (c/d)                & (ppt)         &($^{\circ}$)&($^{\circ}$) &   (c/d)       &   ($^{\circ}$)  &      (c/d)      & ($^{\circ}$)   & ($^{\circ}$)  \\
     \hline
f$_1$ &20.1621 $\pm$ 0.0023 &0.95 $\pm$ 0.03 &241 $\pm$ 1&[170:246]& 20.1636 $\pm$ 0.0007 & 243 $\pm$  8 & 20.1620 $\pm$ 0.0012 & 258 $\pm$ 12 & 15 $\pm$ 15 \\
f$_2$ &21.0606 $\pm$ 0.0023 &0.93 $\pm$ 0.03 &122 $\pm$ 1&[21:185] & 21.0616 $\pm$ 0.0007 & 171 $\pm$  7 & 21.0577 $\pm$ 0.0089 & 157 $\pm$ 13 &-14 $\pm$ 15 \\
f$_3$ &~9.8436 $\pm$ 0.0023 &0.79 $\pm$ 0.03 &195 $\pm$ 1&[82:258] & ~9.8445 $\pm$ 0.0009 & 249 $\pm$  7 & 9.8428 $\pm$ 0.0016 & 274 $\pm$ 11 & 25 $\pm$ 13 \\
f$_4$ &24.8835 $\pm$ 0.0017 &0.42 $\pm$ 0.03 &115 $\pm$ 2&[54:126] & 24.8835 $\pm$ 0.0004 & 118 $\pm$  7 & 24.8835 $\pm$ 0.0009 & 115 $\pm$ 11 & -3 $\pm$ 13 \\
f$_5$ &20.5353 $\pm$ 0.0013 &0.71 $\pm$ 0.03 &212 $\pm$ 1&[115:275]& 20.5329 $\pm$ 0.0010 & 256 $\pm$ 10 & 20.5356 $\pm$ 0.0011 & 274 $\pm$ 12 & 18 $\pm$ 15 \\
f$_6$ &34.1252 $\pm$ 0.0027 &0.49 $\pm$ 0.03 &218 $\pm$ 2&[152:230]& 34.1227 $\pm$ 0.0004 & 223 $\pm$  7 & 34.1268 $\pm$ 0.0009 & 230 $\pm$  8 &  7 $\pm$ 11 \\
f$_7$ &~8.3084 $\pm$ 0.0025 &0.63 $\pm$ 0.03 &194 $\pm$ 2&[42:204] & ~8.3107 $\pm$ 0.0010 & 188 $\pm$ 10 & 8.3062 $\pm$ 0.0012 & 202 $\pm$ 16 & 14 $\pm$ 19 \\
f$_8$ &10.8249 $\pm$ 0.0030 &0.64 $\pm$ 0.03 & 11 $\pm$ 2&[10:196] & 10.8236 $\pm$ 0.0010 & 110 $\pm$  8 & 10.8256 $\pm$ 0.0061 & 103 $\pm$ 13 & -7 $\pm$ 15 \\
     \hline
  \end{tabular}
  }
\end{table*}

\begin{figure}[ht!]
  \centering
  \includegraphics[width=.5\textwidth]{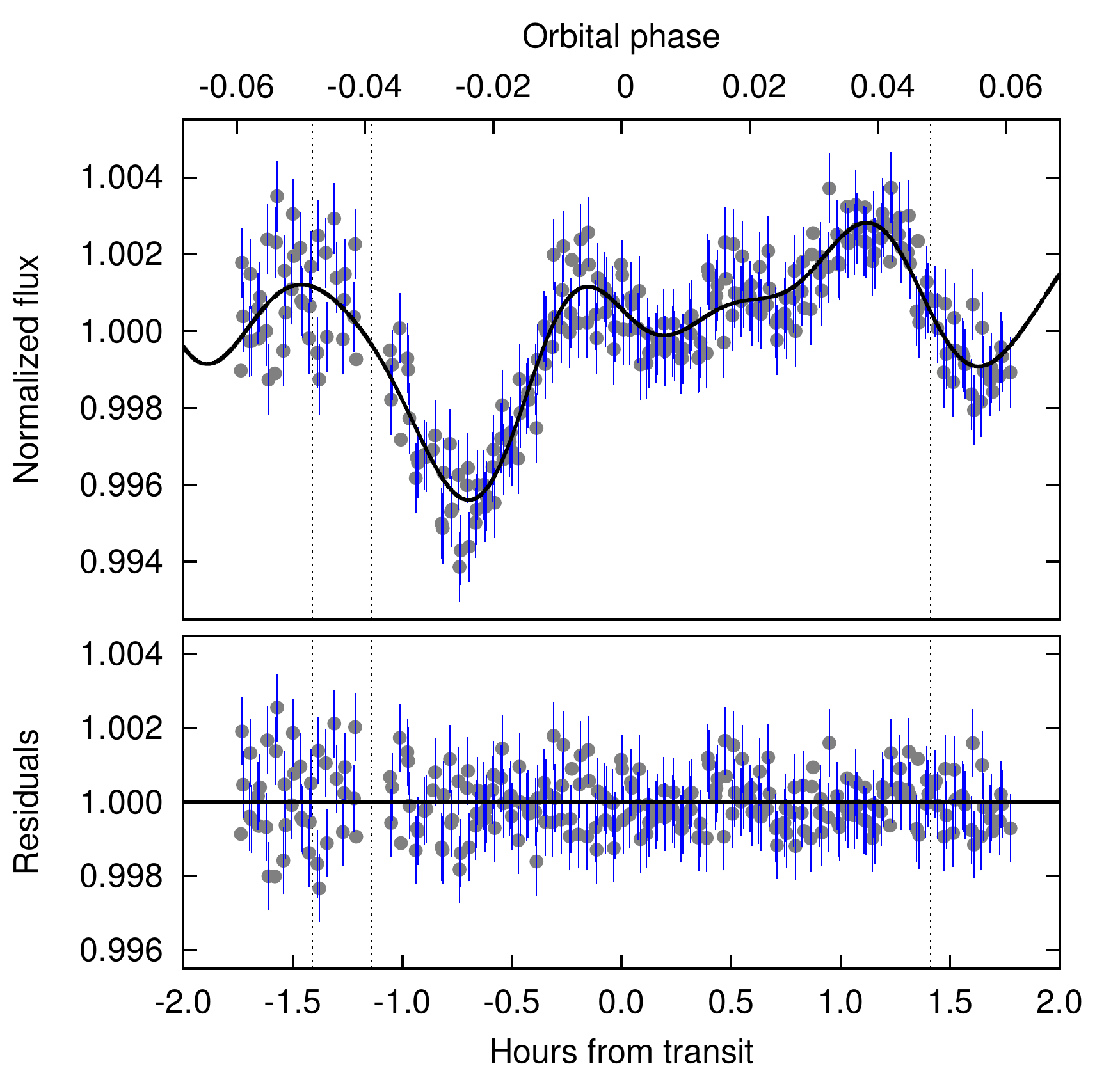}
  \caption{\label{fig:LBT_V} LBT $V$ optical data during secondary
    eclipse of \waspb\ as a function of the orbital phase and the hours
    relative to mid-eclipse. {\it Top:} Normalized flux. The
    continuous black line shows the best-fitting model (24+3
    parameters). The error bars were enlarged to meet the scatter of
    the data. {\it Bottom:} residual light curve. Vertical dashed
    lines indicate first to fourth contacts assuming a circular
    orbit \citep{Deming2012}.}
\end{figure}

For the fitting procedure we used the photometric errors that IRAF
provides, but enlarged to meet the standard deviation of the data. Our
model to the $V$-band light curve requires three additional free
parameters to account for airmass and color-dependent
extinction. Since the pulsations impede any visual inspection of how
these effects impact on the photometry, the number of parameters was
determined through $\chi^2_{\rm red}$ minimization. To derive the 16+3
parameters for our model we used a Markov-Chain Monte Carlo (MCMC)
approach. Our MCMC calculations make extensive use of
PyAstronomy\footnote{http://www.hs.uni-hamburg.de/DE/Ins/Per/Czesla/
  PyA/PyA/index.html}, a collection of Python routines providing a
convenient interface for fitting and sampling algorithms implemented
in the PyMC \citep{Patil2010} and SciPy \citep{Jones2001}
packages. After \mbox{1$\times$10$^6$} MCMC iterations and a burn-in
of the initial \mbox{2$\times$10$^5$} samples, we computed the mean
and standard deviation \mbox{(1-$\sigma$)} of the posterior
distributions of the parameters and used them as best-fit values and
uncertainties, respectively. Our results for the frequencies and
phases are summarized in the sixth and seventh columns of
Table~\ref{tab:MODEL_PARAMS}. The small errors on the fitted
frequencies are caused due to the priors imposed and due to the
reduction of the fitting parameters compared to the ones in
\cite{vonEssen2014}. In this work the eight amplitudes are considered
as fixed parameters, making as consequence the statistical errors to
appear smaller. To test if the MCMC chains were well-mixed and
converged we followed two approaches. Firstly, we started the MCMC
chains using different initial values. Secondly, we divided complete
chains into sub-chains (normally dividing the \mbox{1$\times$10$^6$}
iterations into 4) confirming afterward that the best-fitting values
obtained from the posterior distributions of each sub-chains were
consistent to each other. Finally, we confirmed convergence by
visually inspecting the traces and histograms of all the fitting
parameters. Figure~\ref{fig:LBT_V} shows the LBT optical data
over-plotted to the best-fit model (top) along with the residual light
curve (bottom).

Our analysis over the $V$-band data was concluded trying to estimate
an upper limit for the eclipse depth. For this end, we added to our
model budget a secondary eclipse shape (see Section~\ref{sec:NIR} for
details on secondary eclipse modeling). After similar iterations and
sample burn-in, we found the best-fit eclipse depth to be
\mbox{$\Delta$F = 0.26 $\pm$ 0.48 ppt}. $\chi^2_{\rm red}$
computations between the two models favor the case where the secondary
eclipse is not considered. As previously assumed, the derived
uncertainty is too large to compute a reasonable \mbox{3-$\sigma$}
upper limit.

\subsubsection{Residual light curve analysis}
\label{sec:V_RLC}

Pulsations not accounted for in our model would manifest themselves as
correlated noise. To quantify to which extent the residual light curve
is affected by it, we followed a similar approach as
\cite{Gillon2006,Winn2008}, and \cite{Carter2009}. We started dividing
the residual light curve into M bins of equal duration. Since our data
are not equally spaced, we calculated a mean value N of data points
per bin. If the data are affected by correlated noise, the sample
standard deviation of the binned data, $\sigma_{\rm N}$, would differ
by a factor $\beta_{\rm N}$ from its theoretical expectation
\citep[see e.g.,][for an extended mathematical
  description]{vonEssen2013}. For data sets free of correlated noise,
$\beta_{\rm N}$ = 1 is expected. $\beta_{\rm N}$'s smaller than 1
might occur due to statistical fluctuations and were neglected in our
analysis. We estimated the amount of correlated noise, $\beta$,
averaging $\beta_{\rm N}$'s that were calculated within bins of 0.5,
0.75, 1, 1.25 and 1.5 times the ingress time. Its derived value,
\mbox{$\beta$ = 1.05}, allows us to neglect the contribution of
correlated noise in our future analysis. Finally, we searched for any
residual pulsations computing a Lomb-Scargle periodogram
\citep{Lomb,Scargle,LombScargle} from LBT-$V$'s residual data and
found no significant peaks. Thus, our pulsation model satisfactory
represents the intrinsic variability of \wasp.

\subsection{Analysis of the NIR light curve}

\subsubsection{Wavelength-dependency of fitting parameters for $\delta$ Scuti stars}
\label{sec:WAVE}

\begin{figure}[ht!]
  \centering
  \includegraphics[width=.5\textwidth]{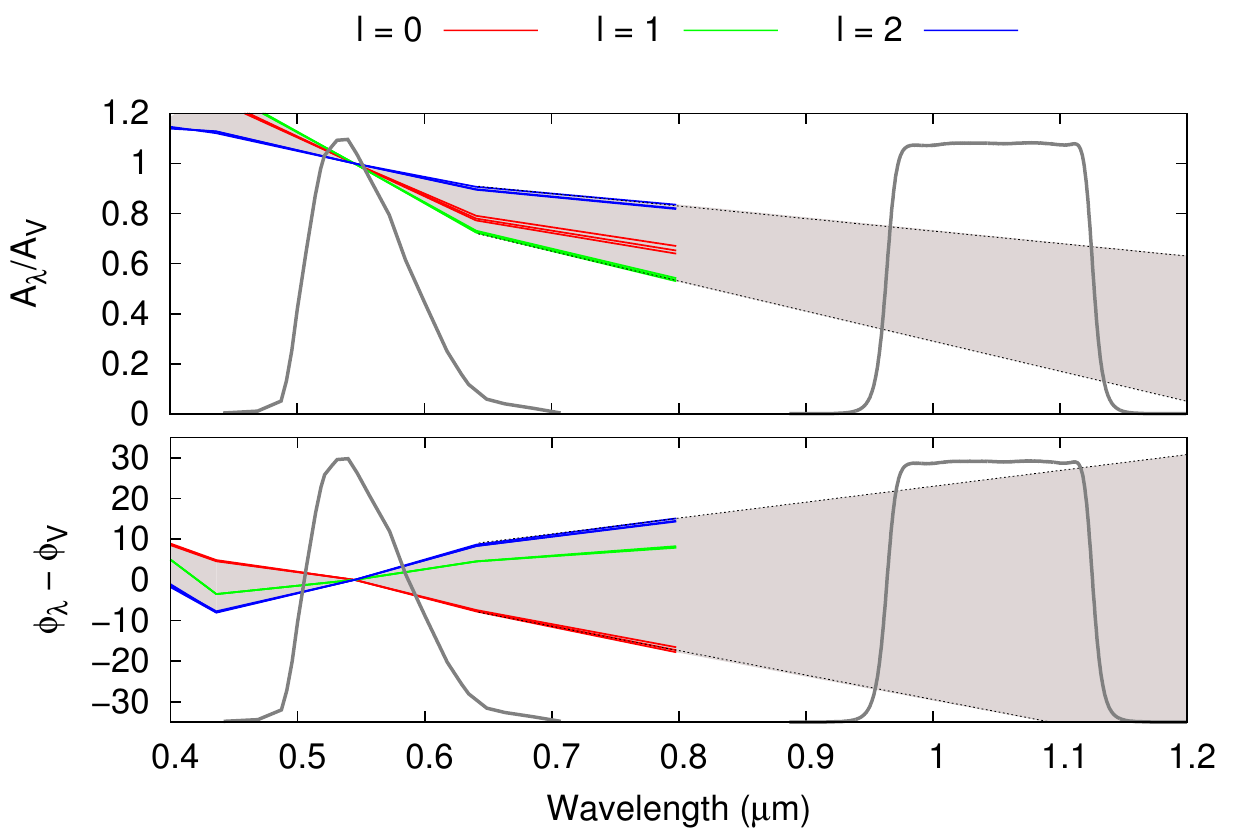}
  \caption{\label{fig:MODES} Amplitude ratios ({\it top}) and phase
    differences {\it bottom} obtained with FAMIAS for the
    \mbox{$\nu_{\rm 1}$ = 20.1621 c/d} pulsation frequency, considering
    \mbox{$\ell\leq$ 2}. Gray contours indicate the extrapolated area
    extended to the $Y$ filter. Both $V$ and $Y$ filter transmission
    functions are plotted in gray continuous lines.}
\end{figure}

The pulsations of $\delta$ Scuti stars are known to be
wavelength-dependent \citep[see
  e.g.,][]{Balona1999,DaszynskaDaszkiewicz2003}. Thus, subtracting the
$V$-model to the $Y$ data would not properly account for \wasp's
intrinsic variability. In \cite{vonEssen2014} we tried to characterize
the nature of the modes analyzing simultaneous multi-color photometry
(see Section 3.6.3 of \cite{vonEssen2014} for further
details). However, we found no conclusive results. Consequently, an
estimation of the $V$--to--$Y$ change of the amplitudes and phases of
the eight frequencies can only be obtained via theoretical
modeling. To quantify these changes we used the Frequency Analysis and
Mode Identification for AsteroSeismology code (FAMIAS), a collection
of tools for the analysis of photometric and spectroscopic time series
data \citep{Zima2008}. To begin with, FAMIAS requires basic stellar
parameters. Therefore, we adopted \mbox{T$_{\rm eff}$ = 7430 $\pm$ 100
  K}, \mbox{log(g) = 4.3 $\pm$ 0.2}, and \mbox{[Fe/H] = 0.1 $\pm$ 0.2}
\citep{CollierCameron2010}. The non-adiabatic observables were
calculated using the code MAD \citep{Dupret2001,Grigahcene2005}, while
as model atmospheres we used \cite{Kurucz1997}'s grids. Furthermore,
we assumed the degree, $\ell$, and the azimuthal number, $m$, of the
pulsation modes to be low \citep[$m,\ell\leq$ 2, see e.g.][geometric
  cancellation]{Aerts2010}. From FAMIAS we computed the amplitude
ratios and phase differences for each one of the eight frequencies
relative to the $V$-band. Unfortunately, the wavelength coverage of
FAMIAS goes up to \mbox{0.8 $\mu$m}, i.e. not covering the $Y$
wavelength range. Therefore, our estimates are an {\it extrapolation}
of FAMIAS results, all of them summarized in
Table~\ref{tab:AMP_PHASE}. As an example, Figure~\ref{fig:MODES} shows
FAMIAS results for \mbox{$\nu_{\rm 1}$ = 20.1621 c/d}, along with
their respective extrapolations. Analyzing these theoretical
expectations some relevant conclusions can be drawn. First, as it is
expected for $\delta$ Scuti stars the amplitudes of the pulsations are
larger where the star emits most of its flux \citep[see
  e.g.,][]{DaszynskaDaszkiewicz2008}. Therefore, for \wasp\ it is
expected that the amplitudes will decrease for redder
wavelengths. Second, the computed amplitude ratios for different
$\ell$ degrees would make the proper identification of the modes
impossible to achieve by means of these multi-color data. Indeed, for
the eight frequencies that comprise our model the amplitudes are
expected to scale down around 40-80\% in the $Y$ with respect to their
$V$ values. Considering the relative short duration of these
observations, the photometric precision of the data, and the relation
between this precision and the amplitude of the pulsations, for low
$\ell$-values it is impossible to determine their geometry (see
Figure~\ref{fig:MODES}). Finally, the phases in $V$ and $Y$ are
expected to be similar. All these aspects will be considered in our
NIR pulsation model.

\begin{table}[ht!]
  \caption{\label{tab:AMP_PHASE} Maximum theoretically-expected
    changes for amplitude ratios
    \mbox{(${A_V/A_Y}_{min}-{A_V/A_Y}_{max}$)} and phase differences
    \mbox{($\phi_{V,Y,min}$,$\phi_{V,Y,max}$)} between the $V$ and
    $Y$-bands for the eight pulsation frequencies considering
    \mbox{$\ell\leq$ 2}.}

  \centering
  \begin{tabular}{c c c c}
    \hline \hline
    PN        & Frequency (c/d)&   $\Delta$A$_{max}$  &   $\Delta\phi_{max}$  ($^{\circ}$,$^{\circ}$)   \\
    \hline
     f$_1$ & 20.1621        & 0.3-0.7    &  (-30,20)  \\
     f$_2$ & 21.0606        & 0.6-0.9    &  (-20,25)  \\
     f$_3$ &  9.8436        & 0.5-0.8    &  (-23,20)  \\
     f$_4$ & 24.8835        & 0.5-0.8    &  (-20,20)  \\
     f$_5$ & 20.5353        & 0.4-0.75   &  (-25,20)  \\
     f$_6$ & 34.1252        & 0.5-0.7    &  (-20,20)  \\
     f$_7$ &  8.3084        & 0.45-0.75  &  (-30,20)  \\
     f$_8$ & 10.8249        & 0.35-0.75  &  (-25,20)  \\
     \hline
  \end{tabular}
\end{table}

\subsubsection{Planetary size versus secondary eclipse model}
\label{sec:NIR}

As secondary eclipse model we used a simple step function. Its shape
was limited by the first to fourth contacts \citep{Sackett1999}. The
latter were estimated using the semi-major axis $a/R_{\rm s}$, the
orbital inclination $i$, the planet-to-star radius ratio $R_{\rm
  P}/R_{\rm S}$, the orbital period $P$, and the mid-eclipse time
$T_{\rm 0}$ that were computed in \cite{vonEssen2014} accounting for
the stellar pulsations. $T_{\rm 0}$ was used to place the step
function relative to mid-eclipse time, assuming a circular orbit
\citep{Deming2012}. The chosen values are summarized in
Table~\ref{tab:ORB_PARAMS}.

\begin{table}[ht!]
  \caption{\label{tab:ORB_PARAMS} Orbital and physical parameters of
    \waspb\ computed in \cite{vonEssen2014}.}
  \centering
  \begin{tabular}{c c}
    \hline \hline
    Orbital parameters & Best-fit values and 1-$\sigma$ errors \\
                       & \citep{vonEssen2014} \\
    \hline
    P (days)           & 1.2198675 $\pm$ 11$\times$10$^{-6}$ \\
    $T_{\rm 0}$ (BJD$_{\rm TBD}$) & 2455507.5222 $\pm$ 0.0003 \\
    $a/R_{\rm S}$            & 3.68  $\pm$ 0.03\\
    i ($\deg$)         & 87.90 $\pm$ 0.93 \\
    $R_{\rm P}/R_{\rm S}$           & 0.1046 $\pm$ 0.0006 \\
    \hline
  \end{tabular}
\end{table}

Any potential change in the planetary radius as function of observing
wavelength mainly manifests itself in a change of the eclipse
shape. In addition, the times of contact might change by a small
amount. We tested if our data is sensitive to this second order effect
and if the use of one set of contact points for all wavelengths is
justified. We simulated a secondary eclipse of a transiting planet
20\% larger than the value measured in the optical
(Table~\ref{tab:ORB_PARAMS}), adding to it the time stamps and noise
properties of our data. We find that a change in the eclipse duration
does not improve the fit to the simulated data compared to a fit where
the radius obtained in the optical is used. Our data quality is such
that we do not need to take this second order effect into account and
we proceed using the times of contact as determined in the optical.

\subsection{Fitting parameters and results}

We now use our knowledge on the pulsations gained from the analysis of
the optical light curve in the analysis of the NIR light curve. Our
model consists of an eclipse light curve described by the parameters
from Table~\ref{tab:ORB_PARAMS}. For all but the planet to star flux
ratio \mbox{$\Delta$F}, which we want to determine, we use the
parameters and uncertainties given in the table as prior
information. To describe the stellar pulsations we use the same model
as in the optical plus the prior information on the eight frequencies,
amplitudes and phases. All amplitudes are damped by a common factor
\mbox{$\Delta$A} relative to the optical. All parameters and their
1-$\sigma$ uncertainties are given in Table~\ref{tab:MODEL_PARAMS}. In
particular, we find an eclipse depth of \mbox{$\Delta {\rm F} = 1.03
  \pm 0.34$ ppt} in the $Y$-band.

\begin{figure}[ht!]
  \centering
  \includegraphics[width=.5\textwidth]{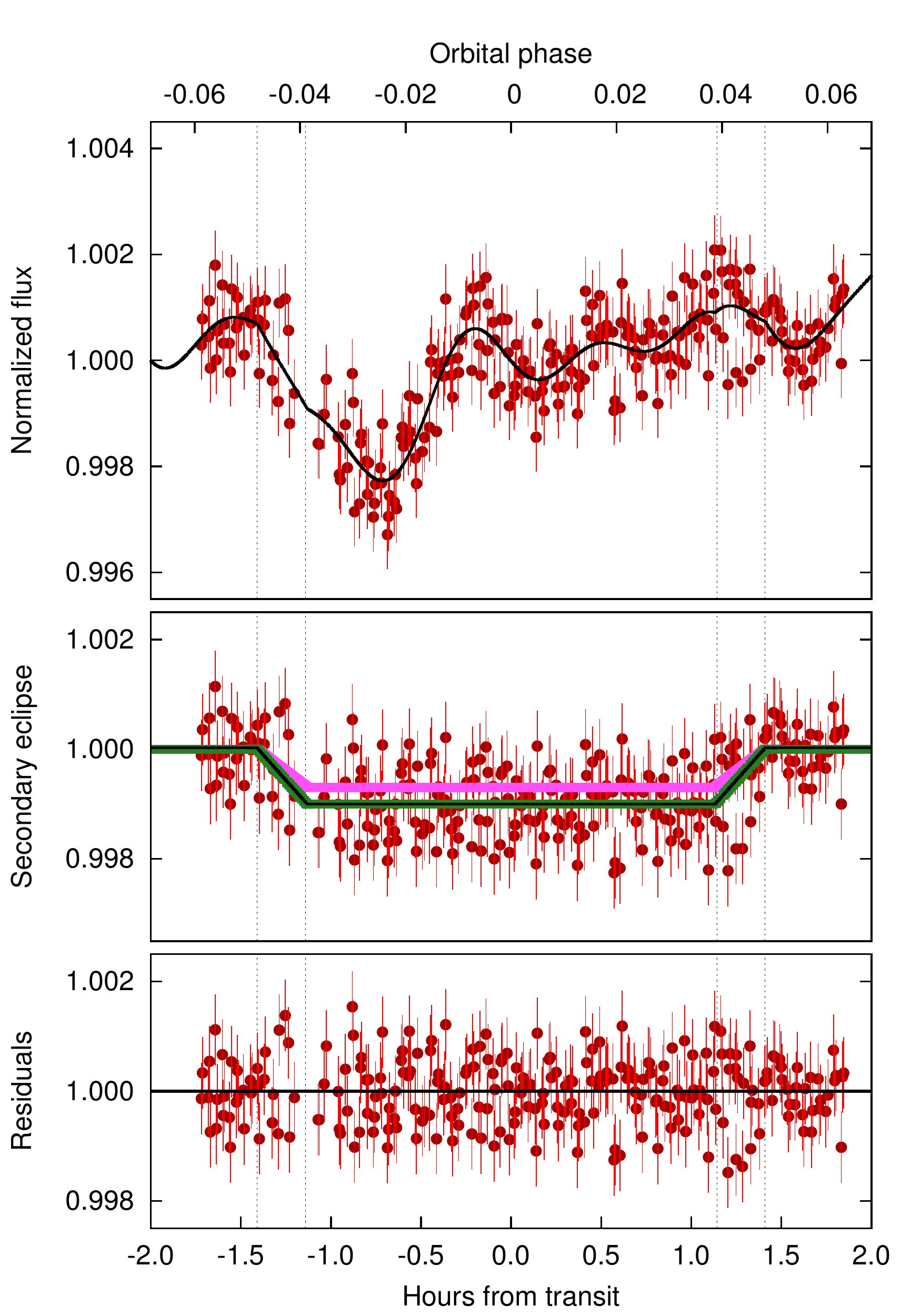}
  \caption{\label{fig:LBT_Y}Equivalently to Figure~\ref{fig:LBT_V},
    but around 1.05 $\mu$m. From top to bottom, \wasp's pulsations and
    secondary eclipse, the secondary eclipse once the pulsations have
    been removed, and the residual light curve once the secondary
    eclipse is removed. Black continuous lines show best-fit models
    for the pulsations and the secondary eclipse. Pink and green
    eclipse models correspond to \cite{Deming2012}'s predicted eclipse
    depths for inverted and non-inverted atmospheric structures
    integrated in the $Y$-band.}
\end{figure}

The derived frequencies, phases, and the computed phase differences
between both bands, are summarized in the last three columns of
Table~\ref{tab:MODEL_PARAMS}. Our results show that both $\phi_{\rm
  V}-\phi_{\rm Y}$ and \mbox{$\Delta$A = 0.73 $\pm$ 0.08} are
consistent with their theoretical expectations. Figure~\ref{fig:LBT_Y}
shows in the top panel the $Y$-band light curve, in the middle panel
the pulsation model removed isolating the planetary eclipse and in the
bottom panel the residuals model data. On top of our best-fit
secondary eclipse model we show \cite{Deming2012}'s predicted eclipse
depths integrated in the $Y$-band for the atmosphere of \waspb\ when a
carbon-rich non-inverted model (green) and a solar composition model
with an inverted temperature structure (pink) are considered.

We calculated the Pearson, $P$, and Spearman, $S$, correlation
coefficients between \mbox{$\Delta$A} and \mbox{$\Delta$F}. Both show
a low positive correlation of \mbox{$\sim$0.4}. No such correlation
was found between the eclipse depth and the phases nor frequencies.

\subsection{Robustness of our fitting approach}

Measured eclipse depths and uncertainties can depend sensitively on
the choice of model used to fit the light curve. To confirm that our
final result is robust, we performed a number of additional fits. We
compared the models using the Bayesian information criterion,
\mbox{BIC = $\chi$ + k ln N}, which penalizes the number k of model
parameters given N data points. The additional models for the stellar
pulsations included the following fitting parameters:

\begin{enumerate}
\item Frequencies fixed to the best-fit values found in
  \cite{vonEssen2014}, scaling factor for the amplitudes, eight phases
  with Gaussian priors, secondary eclipse and detrending function.
\item Eight frequencies with Gaussian priors, eight amplitudes with
  uniform priors, eight phases with Gaussian priors, secondary eclipse
  and detrending function.
\item Frequencies fixed to the best-fit values found in
  \cite{vonEssen2014}, eight amplitudes with uniform priors, eight
  phases with Gaussian priors, secondary eclipse and detrending
  function.
\end{enumerate}

The resulting BIC values, along with the fitting parameters for each
model and the derived best-fit eclipse depth for each model are
summarized in Table~\ref{tab:BIC_vs_ED}. As the table reveals, the BIC
favors our approach. All eclipse depths are consistent within
1-$\sigma$ errors.

\begin{table}[ht!]
  \caption{\label{tab:BIC_vs_ED} Results for the four different
    pulsation models. From left to right, stellar pulsation model
    (Model 0 corresponds to the one carried out originally, while the
    remaining numeration follows the previously detailed stellar
    models), the number of fitting parameters, k, the computed
    $\chi^2$ value between the observations and the best fitting
    models, the Bayesian Information Criterion (BIC) and the derived
    eclipse depth (ED).}  \centering
  \begin{tabular}{c c c c c}
    \hline \hline
    Model   &   k    &   $\chi^2$   &   BIC   & ED (ppt) \\
    \hline
    0       &  21    &  25.3        &  141.1  & 1.03 $\pm$ 0.34  \\
    1       &  13    &  243.4       &  315.1  & 1.23 $\pm$ 0.24  \\
    2       &  28    &  27.3        &  181.7  & 0.98 $\pm$ 0.45  \\
    3       &  20    &  249.1       &  359.4  & 1.24 $\pm$ 0.25  \\
    \hline
  \end{tabular}
\end{table}

\subsection{Potential third-light contributions}
\label{lab:thirdLight}

\cite{Moya2011} detected an optical companion to \wasp\ at an angular
separation of \mbox{$\sim$2''}. This companion is included in our
aperture and therefore reduces the measured eclipse depth relative to
the undiluted case. Based on the color information obtained in the
J$_{\rm C}$, H$_{\rm C}$, K$_{\rm C}$, and FII filters,
\cite{Moya2011} estimated that the probable physical companion of
\wasp\ has an effective temperature of\mbox{ $T_{\rm eff}$ = 3050
  $\pm$ 250 K}. To compute the contribution of this third light we
used the approach detailed in \cite{Tingley2004}. First, we
represented the emission of the third body, $F_{\rm 3}$, relative to
\wasp's emission, F$_{\rm W33}$, with the ratio of two black body
curves \mbox{F$_{\rm 3}$/F$_{\rm W33}$} of 3050 and 7430 K,
respectively. Then, we converted the \mbox{$\Delta$mag's} of the
J$_{\rm C}$, H$_{\rm C}$, K$_{\rm C}$, and FII bands to fluxes and
adjusted F$_{\rm 3}$/F$_{\rm W33}$ to \cite{Moya2011}'s measurements
with a diluting (scaling) factor, $C$, that accounts for the relative
distances between the two stellar objects. Extrapolating this to the
$Y$-band we found F$_{\rm 3}$ to be 714 times dimmer than \wasp. Given
our data we can ignore such a contribution. Although strictly
  speaking stars are not blackbodies, the blackbody approximation was
  carried out in this case mainly due to two reasons. On one hand,
  \wasp\ is an A-type star and, therefore, is expected to have no
  strong absorption lines in the wavelength range of our interest. On
  the other hand, there is no knowledge about the metallicity nor the
  surface gravity of \wasp's companion, not allowing us to use
  appropriate stellar models without further speculation. Nonetheless,
  we caution the reader to carry out this approximation when a more
  complete knowledge of the spectral properties of the involved stars
  is known.

\section{Thermal emission from WASP-33}
\label{SecTrans}

\subsection{Brightness and equilibrium temperatures}

From the observed planet-star flux ratio at a given wavelength the
exoplanet brightness temperature $T_{\rm b}$ at that wavelength can be
derived. For this calculation we represent \wasp's emission by a
spectrum from the PHOENIX library \citep{Peter1,Peter2} with stellar
parameters \mbox{T$_{\rm eff}$ = 7400 K}, \mbox{[Fe/H] = 0.1}, and
\mbox{log g = 4.5}, closely matching \wasp's values given by
\cite{CollierCameron2010}. The planet contribution is represented by a
black body and the measured eclipse depth represents the
planet-to-star flux ratio \mbox{F$_{\rm P}$/F$_{\rm S}$(Y)},
integrated within the $Y$-band \citep{Charbonneau2005}. We computed
$\chi^2$ between our best-fit eclipse depth and F$_{\rm P}$/F$_{\rm
  S}$(Y), varying $T_{\rm b}$ between 2500 and 4500 K each \mbox{1
  K}. By means of $\chi^2$ minimization we found the brightness
temperature of the exoplanet to be \mbox{T$_{\rm b}$(Y) = 3398 $\pm$
  302}. Its error was determined from a $\Delta\chi^2$ contour equal
to 1. 

\begin{figure}[ht!]
  \centering
  \includegraphics[width=.43\textwidth]{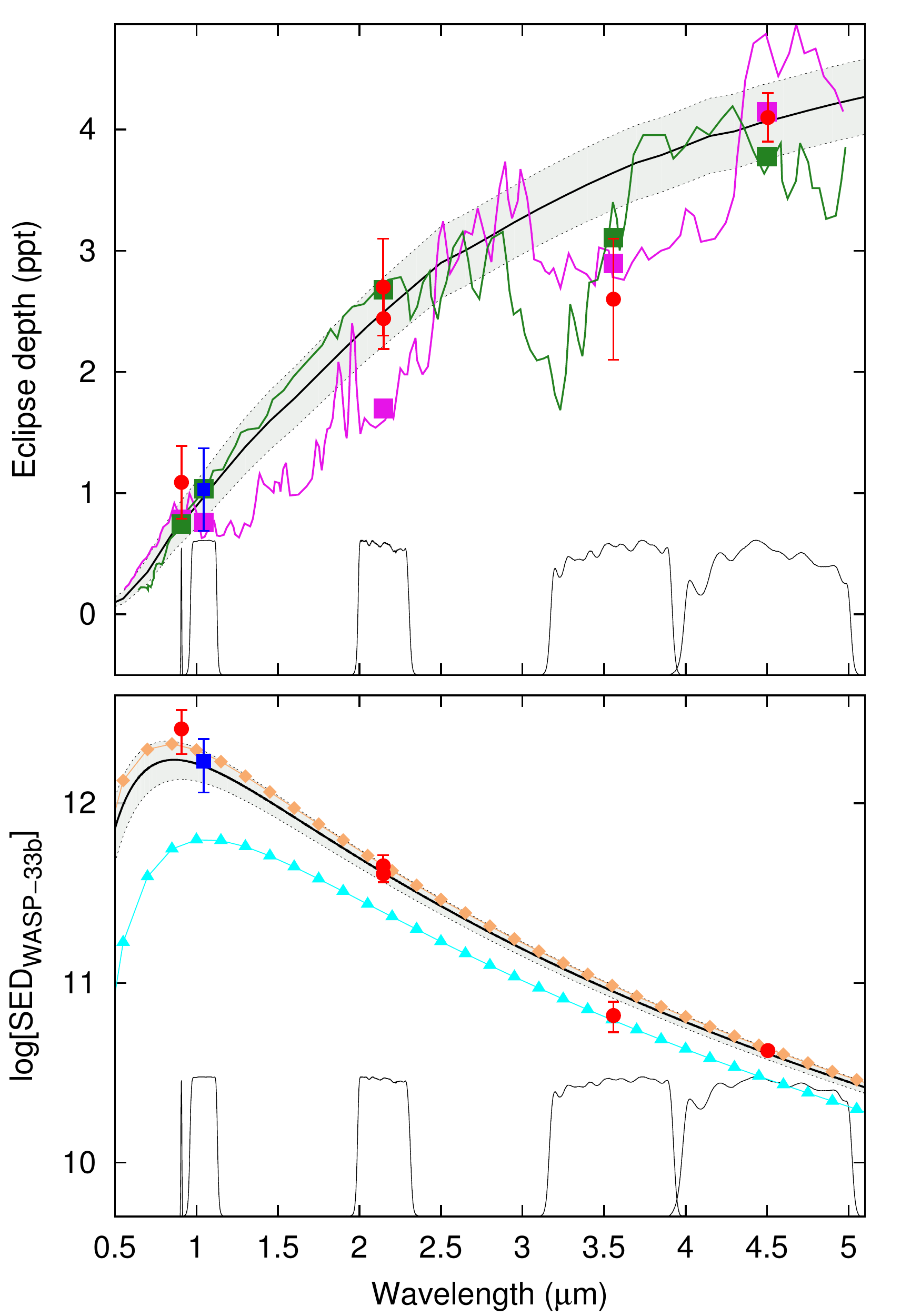}
  \caption{\label{fig:Teff} {\it Top:} Eclipse depth in ppt as a
    function of wavelength. The blue data point indicates our
    brightness temperature measurement, while the red circles show
    literature values. The gray area shows 1-$\sigma$ contour of
    \waspb's equilibrium temperature. Green continuous line
    corresponds to \cite{Deming2012}'s carbon-rich non-inverted model,
    while pink continuous line indicates a solar composition model
    with an inverted temperature structure. Squares mark the planetary
    fluxes integrated over the bandpasses. {\it Bottom:} The same data
    after multiplication by the PHOENIX stellar model revealing the
    Spectral Energy Distribution (SED) for \waspb. Overplotted are the
    expected SEDs for a zero-albedo instantly re-radiation (orange
    diamonds, f = 2/3) and instantly redistribution (cyan triangles, f
    = 1/4) day-sides.}
\end{figure}

To date, six measurements of \waspb's eclipses along with their
corresponding $T_{\rm b}$'s have been obtained
(Table~\ref{tab:EDs_BIB}). Assuming that the brightness temperatures
can represent the equilibrium temperature T$_{\rm eq}$ of
\waspb\ \citep[see e.g.,][Section 4.1]{deMooij2013}, we re-determined
it following the approach explained in the previous paragraph but
using all bibliographic data instead. We found the best-fit
equilibrium temperature equal to \mbox{T$_{\rm eq}$ = 3358 $\pm$ 165
  K} (Figure~\ref{fig:Teff}). Our determination of the equilibrium
temperature is \mbox{$\sim$100 K} higher than, but consistent with,
the value determined by \cite{deMooij2013}, \mbox{T$_{\rm eq}$ =
  3298$^{+66}_{-67}$ K}. The corresponding $\chi^2_{\rm red}$ at
\mbox{5 d.o.f.}  is 1.2. We calculated $\chi^2_{\rm red}$ values
between the observations and the two cases presented in
\cite{Deming2012} (see their Figure 9 and our Figure~\ref{fig:Teff}),
i.e. a carbon-rich non-inverted model (green, $\chi^2_{\rm red}$ =
1.95, 3 d.o.f.), and a solar composition model with an inverted
temperature structure (pink, $\chi^2_{\rm red}$ = 5.71, 3 d.o.f.). We
find that our simple blackbody curve seems to provide comparable
results to the non-inverted scenario, and could also easily fit the
inverted scenario if the emission from the TiO band centered at 0.9
$\mu$m is stronger than what current models predict. We notice that
the measurement at 2 microns differs significantly with respect to the
inverted (pink) model. This discrepancy has been solved by
\cite{Haynes2015} (see their Section 5.1 and their Figure 5). We
believe that these data can not be uniquely represented by a given
model without the need of further speculation, and find our simple
blackbody approach as sufficient. A more sophisticated analysis using
different models with different inverted temperature structures is
beyond our scope.

Special notice has to be given to the wavelength coverage of the
data. The lower panel of Figure~\ref{fig:Teff} shows that all
planet-to-star flux ratio measurements lie within the Rayleigh-Jeans
region of the planets black body, making any temperature measurement
prone to errors, since the raising part of the planetary emission
curve is not being taken into account. A flux measurement at
\mbox{$\lambda\sim$0.7-0.9 $\mu$m} would significantly improve our
knowledge on \waspb's temperature, because the shape of the planetary
SED would be better constrained and the determination of the planetary
temperature wouldn't be determine only by the slope of the
SED. However, secondary eclipse observations in the optical are
extremely challenging, aggravated in this particular case by \wasp's
intrinsic variability.

\begin{table}[ht!]
  \centering
  \caption{\label{tab:EDs_BIB} Observed eclipse depths (EDs) of
    \waspb\ in ppt and the derived brightness temperatures (T$_{\rm
      b}$). CW corresponds to the central wavelength of the filters.}
  \begin{tabular}{l c c c}
    \hline \hline
    Authors                &    EDs           & T$_{\rm b}$  &   CW          \\
                           &   (ppt)          &  (K)       &   ($\mu$m)     \\ 
    \hline
    \cite{Smith2011}       &  1.09 $\pm$ 0.30 & 3490 $\pm$ 140 &   0.91     \\
    \cite{Deming2012}      &  2.70 $\pm$ 0.40 & 3415 $\pm$ 130 &   2.14     \\
    \cite{Deming2012}      &  2.60 $\pm$ 0.50 & 2740 $\pm$ 225 &   3.55     \\
    \cite{Deming2012}      &  4.10 $\pm$ 0.20 & 3290 $\pm$ 100 &   4.50     \\
    \cite{deMooij2013}     &  2.44 $\pm$ 0.25 & 3270$^{+115}_{-160}$ &   2.14  \\
    This work              &  1.03 $\pm$ 0.34 & 3398 $\pm$ 302 &   1.05     \\
    \hline
  \end{tabular}
\end{table}

\subsection{Albedo and re-distribution factor}

Due to their proximity to the host star, the exoplanet atmospheres of
short period exoplanets are strongly dominated by the stellar incident
flux. An indicator for the fraction of light that gets reflected back
into space immediately is the geometric albedo for which the majority
of measurements yielded very low values \citep{Heng2013}. How well the
absorbed flux is redistributed from the dayside to the nightside is
still a matter of debate
\citep{Perna2012,Heng2014}. \cite{LopezMorales2007} presented an
analytic expression for the exoplanet equilibrium temperature as a
function of the Bond albedo $A_{\rm B}$ and the reradiation factor
$f$, which describes how the stellar radiation absorbed by an
exoplanet is redistributed in its atmosphere. Using their Eq. 1, we
compute the equilibrium temperature of \waspb\ for a zero-albedo and
reradiation factors of 2/3 (instant reradiation) and 1/4 (instant
redistribution) obtaining \mbox{$T_{\rm eq}$ = 3499 K} and
\mbox{$T_{\rm eq}$ = 3257 K}. These are shown in the bottom panel of
Figure~\ref{fig:Teff} as orange diamonds and cyan triangles,
respectively.

\begin{figure}[ht!]
  \centering
  \includegraphics[width=.5\textwidth]{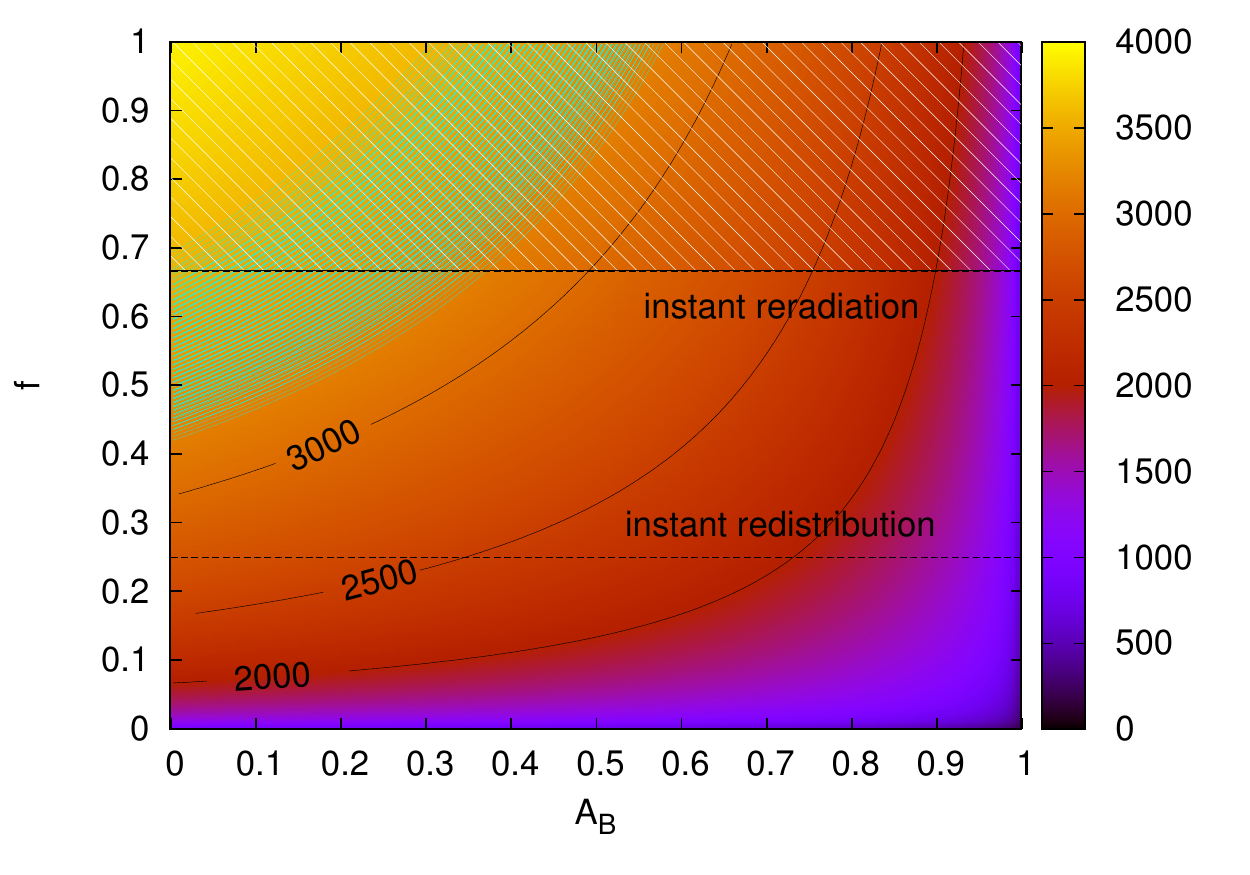}
  \caption{\label{fig:map} Equilibrium temperature map as a function
    of albedo \mbox{($A_{\rm B}$)} and re-distribution factor
    $f$. Cyan color shows the contours relative to \waspb's
    equilibrium temperature within 1-$\sigma$ error (top-left corner)
    for both planetary and stellar parameters. f values larger
      than 2/3 are unphysical. Therefore, the white-shaded area of the
      plot has not to be considered.}
\end{figure}

Instead of choosing two boundary conditions for $A_{\rm B}$ and $f$,
to constrain \waspb's values of bond albedo and redistribution
efficiency we produced an equilibrium temperature map. Each point of
the map was calculated in the same fashion as before, but defining a
\mbox{300$\times$300} grid in the $A_{\rm B}-f$ space. The map is
displayed in Figure~\ref{fig:map}, where the cyan area defines
\waspb's equilibrium temperature range. The shaded effect was produced
considering the stellar \mbox{(T$_{\rm eff}$, R$_{\rm S}$)} and
orbital ($a/R_{\rm S}$) parameters with their uncertainties.

Producing a statistical analysis of the bond albedo and redistribution
efficiency over 24 transiting exoplanets with at least one published
secondary eclipse, \cite{Cowan2011} found that all the planets in the
sample were consistent with low bond albedos ($<10$\%), and planets
hotter than \mbox{2400 K} presented low redistribution efficiency
(i.e., large $f$ values). This is in good agreement with the cyan
contours displayed in Figure~\ref{fig:map}. Thus, \waspb\ seems to
have similar properties as atmospheres of other close-in gas giants
orbiting cooler stars \mbox{(A$_B<$0.38, 0.41$<$f$<$2/3)}.

A nonzero geometric albedo of the planet in the $Y$ band would result
in a contribution of reflected light to the eclipse depth. For
example, the potential impact of a rather high geometric albedo of 0.5
would be $\sim$350 ppt on the eclipse depth, which is on the order of
our 1-$\sigma$ uncertainty. Such contribution of reflected light would
not significantly affect our results of Sect. 4.1. However, any
nonzero albedo would lower the estimated exoplanet equilibrium
temperature.

\subsection{Evidence for temperature inversion?}

\begin{figure}[ht!]
  \centering
  \includegraphics[width=.5\textwidth]{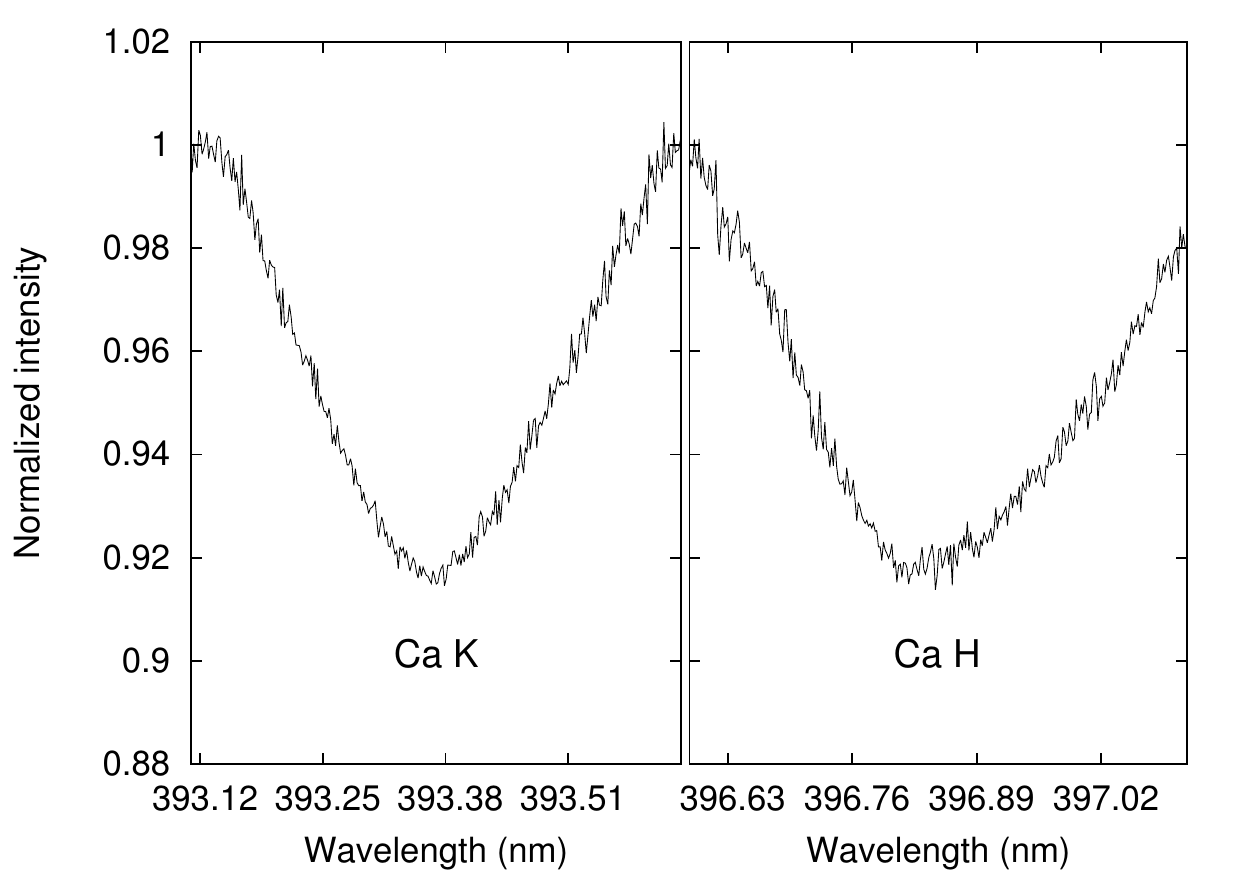}
  \caption{\label{fig:Ca}\wasp's high resolution spectra around the
    Ca II H\&K lines observed with Keck/HIRES.}
\end{figure}

The observed diversity in the characteristics of hot Jupiter
atmospheres has been a main subject of investigation. For instance,
\cite{Knutson2010} suggested a direct relation between the activity of
the host star and the presence of a temperature inversion. While
active stars should host exoplanets without temperature inversions,
the opposite should occur around quiet stars. To investigate \wasp's
activity levels we analyzed calcium lines in high-resolution spectra
acquired on December 28, 2012, using the High Resolution Echelle
Spectrometer mounted at the Keck I telescope (Program ID N116Hr, PI
Howard). \wasp\ is X-ray dark and generally inactive, as seen in a
lack of emission in the Ca II H and K lines
(Figure~\ref{fig:Ca}). Comparing the observations to
\cite{Knutson2010}'s suggestions, \waspb\ should present a temperature
inversion. Also, planets with inversions are expected to have greater
SPITZER/IRAC 4.5 $\mu$m than 3.6 $\mu$m flux, resulting from strong CO
and CO$_2$ absorption in the 4.5 $\mu$m bandpass \citep[see
  e.g.,][]{Burrows2007,Madhusudhan2010}. This has been observed in
\wasp. However, both results seem to be in disagreement with the
observed high re-radiation factor \citep{Fortney2008}. Although
$\chi^2$ analysis would favor a non-inversion, the here presented
disagreements caution ourselves to make any assumption about the
atmospheric structure of \waspb. A phase curve would help to constrain
the albedo and the re-radiation factors. However, stellar pulsations
might make the task more challenging.

\section{Conclusions}
\label{Concl}

In this work we present simultaneous observations in the $V$ and $Y$
bands of \wasp\ during a secondary eclipse. We make use of the
information about the stellar pulsations on the $V$ light curve to
``clean'' the NIR data, minimizing the impact of stellar pulsations
and correlated noise in the determination of the eclipse depth. The
measured eclipse depth is \mbox{$\Delta F$ = 1.03 $\pm$ 0.34 ppt},
which corresponds to a brightness temperature of \mbox{3398 $\pm$ 302
  K}. We represent the spectral energy distribution of \waspb\ by a
black body curve and fit it to current data. We find an equilibrium
temperature equal to \mbox{3358 $\pm$ 165 K}, slightly hotter but
still consistent with previous measurements. Comparing $\chi^2$ values
between our approach and more sophisticated measurements we conclude
that these observations can not provide a detailed description of the
temperature structure of \waspb's atmosphere. In agreement with
previous publications and hypothesis made by \cite{Cowan2011}, we find
that \waspb\ has a very low efficiency to circulate stellar radiation
into the exoplanet night side.

\begin{acknowledgements}

C. von Essen acknowledges D. Deming, A. Mandell, K. Haynes, M. Breger
and B. W. Tingley for fruitful discussions. We thank the LBTO support
astronomers Olga Kuhn, Michelle Edwards, David Thompson and Barry
Rothberg for their immense help in planning and executing the LBT
observations. Funding for the Stellar Astrophysics Centre is provided
by The Danish National Research Foundation (grant No. DNRF106). The
research is supported by the ASTERISK project (ASTERoseismic
Investigations with SONG and Kepler) funded by the European Research
Council (Grant agreement No. 267864). AMSS acknowledges support from
the Polish NCN through grant no. 2012/07/B/ST9/04422. This research
has made use of the Keck Observatory Archive (KOA), which is operated
by the W. M. Keck Observatory and the NASA Exoplanet Science Institute
(NExScI), under contract with the National Aeronautics and Space
Administration.

\end{acknowledgements}

\bibliographystyle{aa}
\bibliography{wasp33}

\end{document}